\documentclass[conference]{IEEEtran}
\IEEEoverridecommandlockouts
\usepackage{cite}
\usepackage{amsmath,amssymb}
\usepackage{bm}
\usepackage{booktabs}
\usepackage{multirow}
\usepackage{array}
\usepackage{graphicx}
\usepackage{url}
\usepackage[T1]{fontenc}
\usepackage{tipa}
\usepackage[usenames,dvipsnames]{color}
\usepackage[hidelinks]{hyperref}

\def\x{{\bm x}}
\def\y{{\bm y}}
\def\z{{\bm z}}
\def\bpi{{\bm \pi}}
\newcommand{\blank}{\langle\mathrm{b}\rangle}

\title{Latent Softmax for Data-Efficient Phoneme-Based Multilingual ASR Across Tonal and Non-Tonal Languages
\thanks{\textsuperscript{*} Corresponding author. This work was supported in part by Guangxi Science and Technology Project (GUIKEAB25069456), and in part by National Natural Science Foundation of China under Grant (62466055).}
\thanks{\textsuperscript{\dag} Work done during internship at TasiTech.}
}

\author{
    \IEEEauthorblockN{Saierdaer Yusuyin$^1$\textsuperscript{\dag}, Nanling Jiang$^2$\textsuperscript{\dag}, Hao Huang$^1$, Zhijian Ou$^{3,4}$\textsuperscript{*}}
    \IEEEauthorblockA{$^1$ School of Computer Science and Technology, Xinjiang University, China}
    \IEEEauthorblockA{$^2$ Department of Electronic Engineering and Information Science, University of Science and Technology of China, China}
    \IEEEauthorblockA{$^3$ SPMI Lab \& IEIT in Tianjin, Tsinghua University, China}
    \IEEEauthorblockA{$^4$ TasiTech Co. Ltd., China}
\IEEEauthorblockA{sar\_dar@foxmail.com, huanghao@xju.edu.cn, ozj@tsinghua.edu.cn}}

\begin{document}
\maketitle

\begin{abstract}
Phoneme-based multilingual automatic speech recognition (ASR) shares acoustic evidence across languages more directly than language-specific subword modeling. Joint training of tonal and non-tonal languages faces a supervision mismatch: tonal languages annotate tone-marked vowels, whereas non-tonal languages typically provide only base-vowel labels. A standard softmax either treats these as unrelated classes, weakening cross-lingual sharing, or collapses tones, losing distinctions required by tonal languages. We propose \emph{Latent Softmax}, a connectionist temporal classification (CTC)-compatible output layer with tone-marked vowels as subclasses and base vowels as major classes; consonants and the CTC blank remain singleton labels. When only a base-vowel label is observed, the tone-marked subclass is latent and marginalized out. Multilingual experiments on AISHELL-1 Mandarin and LibriSpeech English reduce speech-to-phoneme (S2P) phoneme error rates over a standard softmax multilingual baseline by 8.4\%, 17.5\%, and 12.6\% on AISHELL-1, test-clean, and test-other, respectively. These encoder improvements also yield consistent character and word error rate gains with both large-language-model phoneme-to-grapheme conversion (LLM-P2G) and projector-based interfaces. After Mandarin--English code-switching adaptation, Latent Softmax reduces projector-based mixed error rate by 2.6\% on ASRU2019 and 9.5\% on CS-Dialogue; the LLM-P2G results do not establish a consistent advantage.
\end{abstract}


\section{Introduction}

Multilingual and cross-lingual ASR pool data across languages to improve training efficiency, especially in low-resource settings. Recent phoneme-based multilingual ASR uses phonetic supervision as a language-agnostic interface, encouraging finer acoustic-phonetic sharing than language-specific subword units. Reported benefits include multilingual recognition, cross-lingual transfer, and data efficiency \cite{tong2018crosslingual,saier,whistle,yusuyin2025pronunciation}. Weakly phonetic systems such as Whistle further demonstrate the data-efficiency advantage over subword supervision \cite{whistle}. Here, ``data-efficient'' means better use of the same labeled training data through cross-lingual sharing, rather than low-resource scaling with a reduced training set.

For joint tonal/non-tonal training, a straightforward approach places all symbols in one output inventory, treating tone-marked vowels and their non-tonal base vowels as separate units. Mandarin /\=a/ and English /a/, for example, occupy different softmax outputs. This preserves tone distinctions but limits direct sharing: English /a/ frames do not directly provide positive supervision to Mandarin /\=a/, /\'a/, /\v{a}/, and /\`a/, or vice versa. Conversely, collapsing tone-marked variants into one base vowel improves sharing but removes essential tone distinctions. This raises the question: \emph{how can tone modeling enable data-efficient combined training of tonal and non-tonal languages?}

The challenge is to reconcile non-tonal base-vowel major-class labels with the finer tone-marked subclass space required by tonal languages. Non-tonal speech still exhibits pitch variation, but this variation is linguistically non-contrastive and unmarked in phonetic transcription \cite{metze2013models,lin2007switching,lin2006switching}. This motivates a latent view: an English vowel token can instantiate one member of a tone-marked subclass family, even though its subclass identity is unannotated.

\begin{figure*}[t]
  \centering
  \includegraphics[width=0.98\textwidth]{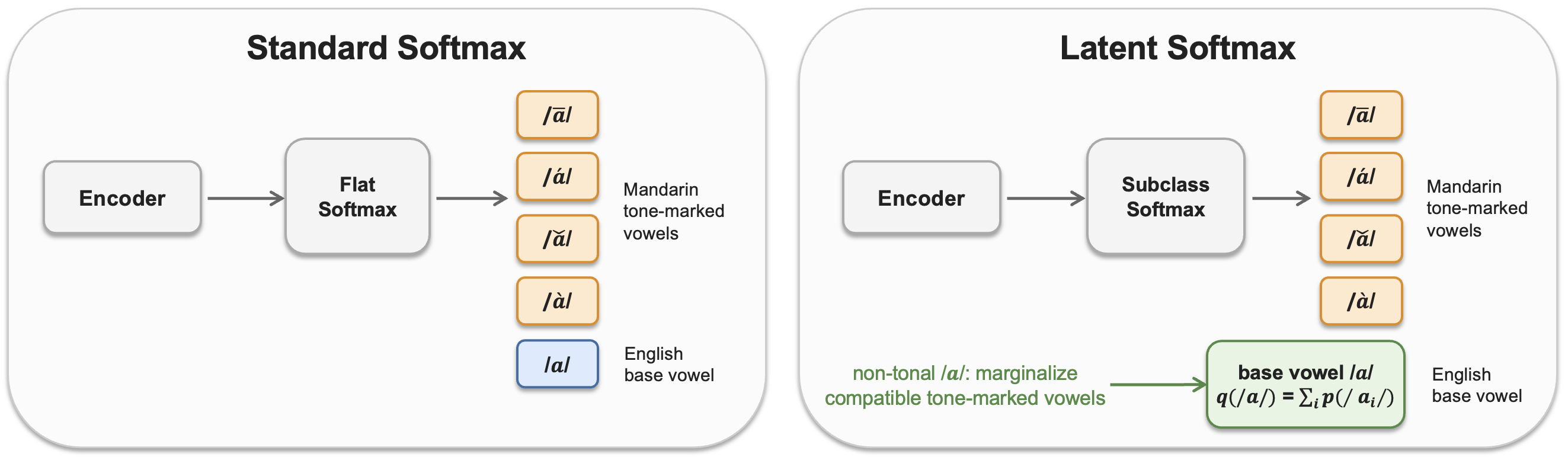}
  \vspace{-2mm}
  \caption{Overview of standard softmax and Latent Softmax for combined training of tonal and non-tonal languages. Standard softmax treats non-tonal base vowels and tone-marked vowels as independent output classes. Latent Softmax keeps tone-marked vowels as subclasses. A tone-labeled sample supervises a singleton tone-marked vowel subclass, while a non-tonal base-vowel sample supervises the base-vowel major class obtained by marginalizing compatible tone-marked vowel subclasses; in this case the subclass identity is latent.}
  \vspace{-4mm}
  \label{fig:latent_softmax_overview}
\end{figure*}

We propose \emph{Latent Softmax}, which computes softmax over subclasses. Tone-marked vowels remain explicit subclasses for tonal languages; when only a base-vowel major-class label is observed, the subclass is latent and marginalized. Non-singleton major classes have no independent output neurons: their probabilities are sums of compatible subclass probabilities. Figure~\ref{fig:latent_softmax_overview} illustrates this mechanism.

This hierarchy enables direct sharing: English /a/ distributes posterior mass across compatible tone-marked subclasses, while Mandarin /\=a/ sharpens one subclass. Non-tonal vowels thus implicitly augment the tone-marked subclass space. Bilingual experiments on AISHELL-1 Mandarin \cite{bu2017aishell} and LibriSpeech English \cite{panayotov2015librispeech} improve S2P and downstream ASR with LLM-P2G \cite{ma2025llm} and projector-based interfaces \cite{li2026phonemesvsprojectorsinvestigation}. In the evaluated Mandarin--English code-switching setting, Latent Softmax improves projector-based MER on both datasets; LLM-P2G shows no consistent advantage.

\section{Related Work}

\subsection{Tone modeling in ASR}

Tone modeling is well studied in tonal-language ASR. Chinese acoustic models commonly use tone-marked vowels, often appending fundamental-frequency (F0) features to frame-level spectral features \cite{lee2002cantonese,lei2006improved}. Recent CTC work compares joint and separate phone-tone modeling: synchronous models with consonants and tone-marked vowels achieve lower error rates on the joint phone+tone tier \cite{li2020autosegmental}. These studies address tonal-language representations; we target mixed tonal/non-tonal training, where non-tonal data lack tone labels.

Tone-related representations can also benefit non-tonal ASR \cite{metze2013models,lin2007switching,lin2006switching}. Latent Softmax promotes cross-lingual sharing by marginalizing tone-marked subclasses under base-vowel supervision, without adding tone features or predicting tone for every language.

\subsection{Multilingual training involving both tonal and non-tonal languages}

Phoneme-based multilingual ASR commonly uses a shared inventory, often based on the International Phonetic Alphabet (IPA), to encourage cross-lingual sharing \cite{tong2018crosslingual,saier,whistle,yusuyin2025pronunciation}. Weakly phonetic systems such as Whistle show that IPA-based supervision can be more data-efficient than subword supervision, but mainly study non-tonal languages and identify tone incorporation as open \cite{whistle}.

Related approaches address different aspects of tonal/non-tonal modeling. JoinAP links phone embeddings through phonological features, evaluating CommonVoice European multilingual training and zero-shot/few-shot transfer to AISHELL-1 Mandarin \cite{zhu2021joinap}. A multilingual allophone system models language-dependent phonemes through language-independent phones \cite{li2020universalphone}. Neither directly addresses the mismatch between tone-marked vowels in tonal supervision and base vowels in non-tonal supervision.

Concatenating tonal and non-tonal vowel inventories preserves distinctions but limits sharing. Latent Softmax instead groups tone-marked subclasses under each base vowel, allowing non-tonal labels to train the subclass space without tone annotations.

\subsection{Relation to hierarchical classification and partial supervision}

Latent Softmax is also related to mixed-granularity and partial supervision. Hierarchical multi-granularity classification accepts coarse or fine labels and maximizes an observed label's marginal probability by aggregating compatible descendants \cite{chen2022label}. Coarse-label representation learning \cite{xu2021coarse} and partial-label learning \cite{lv2020proden} similarly treat supervision as candidate fine-label sets. Our formulation is closest to this marginalization view.

These formulations do not address CTC sequences or multilingual phoneme spaces with tonal and non-tonal supervision. We introduce latent subclass marginalization into this S2P setting to improve cross-lingual data sharing.

\section{Method}

\subsection{Standard softmax for CTC-based S2P}

Let $\mathcal{L}$ contain all CTC alignment labels, including phonemes and the blank, and let $\pi_t$ be the alignment label at frame $t$. Given acoustic input $\mathbf{x}$, standard softmax produces logits $u_t(l)$ for $l\in\mathcal{L}$ and frame posterior
\begin{equation}
 p_t(l)=p(\pi_t=l\mid \mathbf{x})=\frac{\exp(u_t(l))}{\sum_{l'\in\mathcal{L}}\exp(u_t(l'))}.
 \label{eq:std_softmax}
\end{equation}
For target label sequence $\mathbf{y}$, the CTC likelihood is
\begin{equation}
P(\mathbf{y}\mid \mathbf{x}) = \sum_{\pi\in\mathcal{A}(\mathbf{y})} \prod_{t=1}^{T} p_t(\pi_t),
\label{eq:ctc_standard}
\end{equation}
where $\mathcal{A}(\mathbf{y})$ is the set of frame-level alignments that collapse to $\mathbf{y}$.

In joint tonal/non-tonal training, standard softmax treats Mandarin /\=a/ and English /a/ as separate classes or manually collapses them. The former limits sharing; the latter discards tone distinctions. Latent Softmax changes only frame-level emissions, retaining the CTC alignment structure in Eq.~(\ref{eq:ctc_standard}).

\subsection{Latent Softmax formulation}

Let $\mathcal{Z}$ be the subclass inventory over which softmax is computed. It contains tone-marked vowels and singleton labels requiring no subdivision, including consonants and the CTC blank. Each base vowel defines a \emph{major class}; for example, /\=a/, /\'a/, /\v{a}/, and /\`a/ belong to /a/. Singleton labels do not participate in this decomposition.

{\color{black}
Each training utterance comes with a target sequence $\mathbf{y}=(y_1,\ldots,y_N)$. We call each position $y_n$ a \emph{supervised-label occurrence}; its value is a \emph{supervised label}. For a Mandarin utterance, a supervised label may be a tone-marked vowel such as /\=a/. For an English utterance, it may instead be a base vowel such as /a/. In a CTC alignment, $\blank$ is an additional label. For any $l\in\mathcal{L}$, define the compatible-subclass set
\begin{equation}
 B(l)=\{z\in\mathcal{Z}: z\ \text{is compatible with } l\}.
 \label{eq:Bset}
\end{equation}
For a non-tonal base vowel with tone-marked counterparts, $B(l)$ contains all its tone-marked subclasses; otherwise, $B(l)=\{l\}$. The singleton case covers tone-marked vowels, consonants, the blank, and base vowels without tone-marked counterparts (Appendix~\ref{app:inventory}).
}

The network outputs subclass logits $v_t(z)$ and the subclass posterior is
\begin{equation}
 p_t(z)=p(z_t=z\mid\mathbf{x})=\frac{\exp(v_t(z))}{\sum_{z'\in\mathcal{Z}}\exp(v_t(z'))}.
 \label{eq:subclass_softmax}
\end{equation}
{\color{black}
For any fixed supervised label $l$, the quantity
\begin{equation}
 q_t(l)=P\left(z_t\in B(l)\mid\mathbf{x}\right)
 =\sum_{z\in B(l)} p_t(z)
 \label{eq:marginal}
\end{equation}
is an aggregated (marginal) frame posterior, which can be used to define the Latent-Softmax CTC likelihood as shown below.
}
Thus, an English base-vowel label /a/ uses $B(/a/)=\{\text{/\=a/,/\'a/,/\v{a}/,/\`a/}\}$, while a Mandarin tone-marked vowel label /\=a/ uses $B(\text{/\=a/})=\{\text{/\=a/\}}$, assuming four Mandarin tone-marked vowel subclasses. In the phonetic representation used here, tones are attached to the central vowel in a syllable. This is a practical transcription choice: tones are suprasegmental and can extend beyond the vowel segment boundary, but they are conventionally marked on the syllabic nucleus because their most salient effect is typically on the central vowel.

{\color{black}
Appendix~\ref{app:sequence} derives the Latent-Softmax CTC likelihood
\begin{equation}
P(\mathbf{y}\mid \mathbf{x}) = \sum_{\pi\in\mathcal{A}(\mathbf{y})} \prod_{t=1}^{T} q_t(\pi_t),
\label{eq:ctc_latent}
\end{equation}
which has the same form as Eq.~(\ref{eq:ctc_standard}). Each $q_t(l)$ is the marginal probability of a supervised event, not an additional output class; $\sum_{l\in\mathcal{L}}q_t(l)$ need not equal one. Nevertheless, $q_t(l)$ plays the same computational role as $p_t(l)$ in the standard CTC forward--backward recursion and likelihood calculation. In this sense, Latent Softmax is CTC-compatible: a deterministic aggregation layer connects subclass softmax to the standard CTC criterion. Standard forward--backward evaluates Eq.~(\ref{eq:ctc_latent}), and automatic differentiation (e.g., in PyTorch) backpropagates to the network parameters.
}

\begin{figure*}[t]
  \centering
  \includegraphics[width=0.99\textwidth]{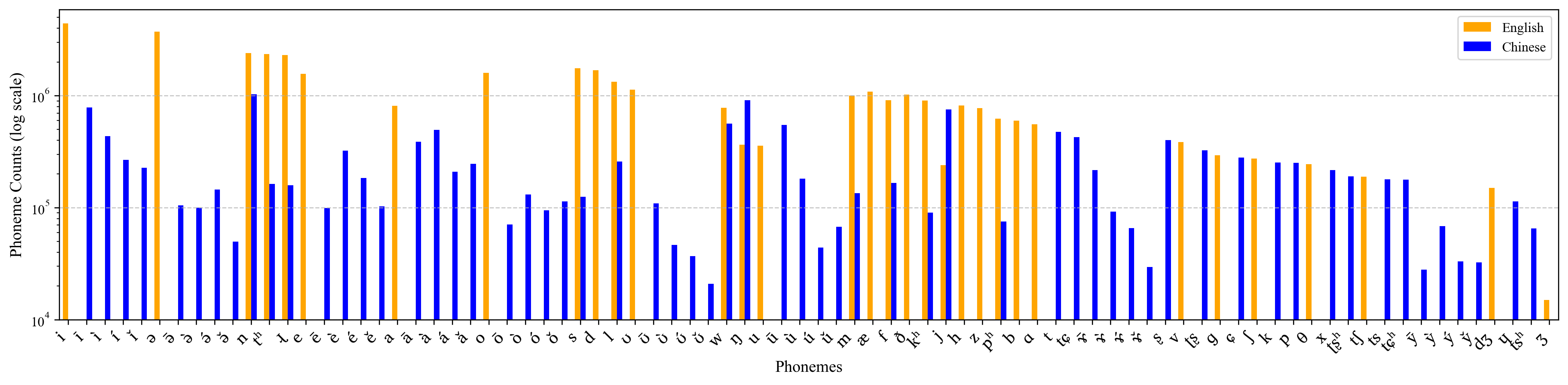}
  \vspace{-2mm}
  \caption{Phoneme-count distribution in LibriSpeech English and AISHELL-1 Mandarin. The horizontal axis lists English and Mandarin phoneme labels, ordered by descending major-class count, and the vertical axis reports their corpus-level counts in the two datasets. When a phoneme appears in both English and Chinese, two bars are shown side-by-side. Base vowels and their tone-marked vowel subclasses, such as /a/, /\=a/, /\'a/, /\v{a}/, and /\`a/, are placed contiguously to show how non-tonal and tonal evidence is distributed across the shared inventory.}
  \vspace{-3mm}
  \label{fig:phoneme_counts}
\end{figure*}

\subsection{Gradient interpretation for Latent Softmax}

For analysis, define the aggregated logit of supervised label $l$ as
\begin{equation}
 u_t(l)=\log \sum_{z'\in B(l)} \exp(v_t(z')).
 \label{eq:agglogit}
\end{equation}
Then the aggregated frame posterior can be rewritten as
\begin{equation}
 q_t(l)=\frac{\exp(u_t(l))}{\sum_{z'\in\mathcal{Z}}\exp(v_t(z'))}.
 \label{eq:qviau}
\end{equation}
Here $u_t(l)$ is only a convenient aggregated logit used for interpretation; Latent Softmax does not introduce separate output neurons for major classes.

We derive log-likelihood gradients with respect to subclass logits; standard backpropagation then gives gradients with respect to network parameters.
{\color{black}
Let $\gamma_t(l)=P(\pi_t=l\mid\mathbf{x},\mathbf{y})$ be the CTC posterior that the frame-level alignment symbol equals $l$ at frame $t$, obtained by the forward--backward algorithm using $q_t(\cdot)$. If the same supervised label $l$ occurs at multiple positions in the target sequence, the forward--backward computation already sums the posterior contributions of all those occurrences into $\gamma_t(l)$. The summation over $l$ below has a different origin: one subclass $z$ can be compatible with multiple supervised labels, so $p_t(z)$ contributes to every aggregated frame posterior $q_t(l)$ for which $z\in B(l)$. Then
}
\begin{align}
\frac{\partial \log P(\mathbf{y}\mid\mathbf{x})}{\partial v_t(z)}
&= \sum_{l} \gamma_t(l)\frac{\partial \log q_t(l)}{\partial v_t(z)} \nonumber\\
&= \sum_{l:z\in B(l)} \gamma_t(l)\frac{\partial u_t(l)}{\partial v_t(z)}-p_t(z) \nonumber\\
&= \sum_{l:z\in B(l)} \gamma_t(l)\frac{p_t(z)}{q_t(l)}-p_t(z).
\label{eq:chainrulegrad}
\end{align}

The same result can also be obtained from a latent-subclass view. Let $z_t$ be the latent subclass at frame $t$, and let $\mathbf{z}=z_1,\ldots,z_T$. By Fisher's identity \cite{ou2018review},
\begin{align}
\frac{\partial}{\partial v_t(z)}\log P(\mathbf{y}\mid\mathbf{x})
&= \mathbb{E}_{P(\mathbf{z}\mid\mathbf{x},\mathbf{y})}
\left[\frac{\partial}{\partial v_t(z)}\log p_t(z_t)\right] \nonumber\\
&= \mathbb{E}_{P(\mathbf{z}\mid\mathbf{x},\mathbf{y})}
\left[\delta(z_t=z)-p_t(z)\right] \nonumber\\
&= P(z_t=z\mid\mathbf{x},\mathbf{y})-p_t(z),
\label{eq:fishergrad}
\end{align}
where the posterior assigned to subclass $z$ is
\begin{align}
&P(z_t=z\mid\mathbf{x},\mathbf{y})\\
=& \sum_{l:z\in B(l)} P(\pi_t=l\mid\mathbf{x},\mathbf{y}) P(z_t=z\mid\mathbf{x},\mathbf{y},\pi_t=l) \nonumber\\
=& \sum_{l:z\in B(l)} \gamma_t(l)\frac{p_t(z)}{q_t(l)}.
\label{eq:posterior_subclass}
\end{align}
Thus Eq.~(\ref{eq:chainrulegrad}) and Eq.~(\ref{eq:fishergrad}) coincide.

The two derivations emphasize complementary aspects of the method. The chain-rule view shows how the positive CTC posterior mass of each supervised label is redistributed across all compatible subclasses. The Fisher-identity view shows the probabilistic meaning: when only a base-vowel label is observed, the tone-marked vowel subclass is latent and its posterior is inferred rather than fixed. Note that for labels other than non-tonal base vowels, $B(l)$ is a singleton and the gradient reduces to the usual CTC-softmax form.

This redistribution lets non-tonal vowel frames update tone-marked subclasses without tone annotations, according to the current model belief. Whereas standard softmax separates English /a/ from Mandarin /\=a/, Latent Softmax distributes English /a/ posterior mass across \{/\=a/,/\'a/,/\v{a}/,/\`a/\}, and Mandarin /\=a/ sharpens one member. The two data sources thus mutually augment each other through the shared base-vowel hierarchy.

\subsection{Evaluation metrics}
We evaluate S2P with tonal phoneme error rate (T-PER) and non-tonal phoneme error rate (NT-PER). Both are edit distances between decoded and reference phoneme sequences, normalized by reference phoneme length. T-PER retains tone marks and is used for AISHELL-1 Mandarin. NT-PER removes tone marks from both sequences and is used for LibriSpeech English and Mandarin--English code-switching, since English subclass outputs may carry tone marks absent from the references. For downstream recognition, we report word error rate (WER) on LibriSpeech and character error rate (CER) on AISHELL-1, using edit distance normalized by reference word and character counts, respectively. For Mandarin--English code-switching evaluation, we report mixed error rate (MER), treating Mandarin characters and English words as mixed tokens and normalizing edit distance by reference token count.

\begin{table*}[t]
  \caption{S2P and downstream recognition error rates (\%). AISHELL-1 uses tonal PER and CER; LibriSpeech uses non-tonal PER and WER. PER is measured at the S2P output.}
  \vspace{-2mm}
  \label{tab:exp}
  \centering
  \begin{tabular}{l|c|c|c|c|c|c}
    \toprule
     \multirow{3}{*}{\textbf{System}} & \multicolumn{2}{c|}{\textbf{AISHELL-1}} & \multicolumn{4}{c}{\textbf{LibriSpeech}} \\
     & \multirow{2}{*}{Tonal PER} & \multirow{2}{*}{CER} & \multicolumn{2}{c|}{test-clean} & \multicolumn{2}{c}{test-other} \\
     &  &  & Non-tonal PER & WER & Non-tonal PER & WER \\
    \midrule
    Standard Softmax S2P + LLM-P2G & \multirow{2}{*}{2.03} & 5.63 & \multirow{2}{*}{1.60} & 4.29 & \multirow{2}{*}{5.00} & 10.70 \\
    Standard Softmax S2P + Projector &  & 5.61 &  & 3.92 &  & 9.61 \\
    \cmidrule{1-7}
    Latent Softmax S2P + LLM-P2G & \multirow{2}{*}{\textbf{1.86}} & \textbf{5.26} & \multirow{2}{*}{\textbf{1.32}} & 3.91 & \multirow{2}{*}{\textbf{4.37}} & 9.89 \\
    Latent Softmax S2P + Projector &  & 5.52 &  & \textbf{3.63} &  & \textbf{9.14} \\
    \bottomrule
  \end{tabular}
  \vspace{-5mm}
\end{table*}

\begin{table}[t]
  \caption{Non-tonal (NT) PERs (\%) and MERs (\%) on Mandarin--English code-switching datasets after code-switching adaptation.}
  \vspace{-2mm}
  \label{tab:code-switch}
  \centering
  \begin{tabular}{l|c|c|c|c}
    \toprule
    \multirow{3}{*}{\textbf{System}} & \multicolumn{2}{c|}{\textbf{ASRU2019}} & \multicolumn{2}{c}{\textbf{CS-Dialogue}} \\
     & NT-  & \multirow{2}{*}{MER} & NT- & \multirow{2}{*}{MER}  \\
      & PER &  & PER &   \\
    \midrule
    Standard Softmax S2P + LLM-P2G & \multirow{2}{*}{\textbf{3.23}} & \textbf{13.69} & \multirow{2}{*}{9.66} & - \\
    Standard Softmax S2P + Projector &  & 14.08 &  & 15.53 \\
    \midrule
    Latent Softmax S2P + LLM-P2G & \multirow{2}{*}{3.83} & 13.83 & \multirow{2}{*}{\textbf{9.02}} & - \\
    Latent Softmax S2P + Projector &  & 13.71 &  & \textbf{14.06} \\
    \bottomrule
  \end{tabular}
  \vspace{-5mm}
\end{table}

\section{Experiments and Results}
\subsection{Datasets and setup}
We conduct multilingual experiments by pooling data from one tonal language (Mandarin) and one non-tonal language (English). 
AISHELL-1 is used as the Mandarin data, containing approximately 178 hours of read speech from 400 speakers \cite{bu2017aishell}. LibriSpeech is used as the English data; it contains approximately 1000 hours of 16~kHz read speech from 2,484 speakers, and we report results on the official test-clean and test-other subsets \cite{panayotov2015librispeech}. 
These datasets were selected primarily because they are widely used open-source corpora for Mandarin and English, respectively. We apply speed perturbation to the AISHELL-1 training data for three-fold data augmentation; LibriSpeech is used without this augmentation. The substantial duration imbalance ($178\times3$ hours versus about 1000 hours) was not introduced as an experimental factor. Although this imbalance may affect the jointly trained multilingual model, it does not confound the controlled comparison between standard softmax and Latent Softmax, because both methods use the same pooled labeled data and training protocol. 
Text transcripts are converted to phoneme sequences using the same lexicon or grapheme-to-phoneme pipeline for all compared S2P systems.

We further evaluate code-switching adaptation on two Mandarin--English code-switching datasets. Both contain utterances in which Mandarin and English may occur within a single utterance. The ASRU2019 Mandarin-English code-switching challenge data contain 500 hours of Mandarin-only speech, a 200-hour Train\_CS set, a 40-hour Dev\_CS set, and a 20.6-hour Test\_CS set \cite{shi2020asru2019}. In our code-switching fine-tuning experiments, the additional 500 hours of Mandarin-only speech are not used. CS-Dialogue is a spontaneous Mandarin-English dialogue corpus with 104.02 hours of speech from 200 speakers and 38,917 utterances \cite{zhou2025csdialogue}. It provides speaker-independent training, development, and test splits of 68.97, 18.30, and 16.74 hours, respectively.

Figure~\ref{fig:phoneme_counts} visualizes the empirical label distribution behind the mixed English--Mandarin training setup. The figure makes explicit the supervision mismatch addressed by Latent Softmax: one base vowel can appear as a non-tonal English label and as multiple Mandarin tone-marked vowel subclasses. Under a standard softmax these counts are consumed by separate output classes, whereas Latent Softmax lets non-tonal base-vowel observations train the corresponding base-vowel major class and, through marginalization, the associated tone-marked vowel subclass family.

The S2P model has 115M parameters. Its Conformer encoder consists of 18 blocks with dimension 512. Each self-attention layer has 8 heads with 64-dimensional hidden states per head, and the feed-forward network (FFN) dimension is 512. We use Qwen3-1.7B \cite{qwen3} as the LLM. Low-Rank Adaptation (LoRA) \cite{hu2022lora} uses rank $r=1024$ and scaling parameter $\alpha=2048$. The projector is a two-layer 4096-dimensional linear projection with a ReLU activation between the two linear layers.

We compare a standard softmax multilingual S2P baseline and the proposed Latent Softmax S2P model. Standard softmax is the most direct controlled baseline: it uses the same S2P encoder, labeled data, and optimization settings, and differs only in output-probability computation---flat labels versus subclass marginalization. Its output layer has 43.1K parameters, compared with 39.5K for Latent Softmax, so Latent Softmax does not benefit from additional output-layer capacity. Tone collapse would remove Mandarin tone distinctions, whereas factorized phone--tone modeling would change both the representation and objective; neither isolates the proposed marginalization mechanism.

Downstream speech recognition is evaluated with two speech-language interfaces: an LLM-P2G system that converts S2P phoneme sequences to text \cite{ma2025llm}, and a projector-based system that maps S2P representations to large-language-model token embeddings \cite{li2026phonemesvsprojectorsinvestigation}.

In the main experiments, the LLM-P2G interface trains the S2P model and the LLM-P2G model in two separate stages. The projector interface adopts a three-stage pipeline: (1) S2P is trained first; (2) after removing the S2P linear classifier, the projector is trained while the S2P encoder and LLM are frozen; (3) finally, the LLM LoRA parameters are trained with both the S2P encoder and projector fixed.

Code--switching adaptation reuses the same stage-wise protocol. All modules are initialized from the main experiment, and the projector is kept frozen throughout. For the projector interface, this means fine-tuning S2P first and then the LLM LoRA parameters. For the LLM-P2G interface, S2P and LLM-P2G are fine-tuned separately in the same order.

\subsection{Main results}

Table~\ref{tab:exp} shows that Latent Softmax reduces S2P PER from 2.03\% to 1.86\% on AISHELL-1, 1.60\% to 1.32\% on LibriSpeech test-clean, and 5.00\% to 4.37\% on test-other. Relative reductions over standard softmax are 8.4\%, 17.5\%, and 12.6\%, respectively. AISHELL-1 uses tonal PER; LibriSpeech uses non-tonal PER.

The downstream error rates show the same trend. With LLM-P2G, Latent Softmax reduces AISHELL-1 CER from 5.63\% to 5.26\% and LibriSpeech WER from 4.29\% to 3.91\% on test-clean and from 10.70\% to 9.89\% on test-other. With the projector, CER decreases from 5.61\% to 5.52\%, and WER from 3.92\% to 3.63\% and from 9.61\% to 9.14\%, respectively. Gains are largest for LLM-P2G, whose input directly depends on decoded S2P phonemes. The projector is already strong on LibriSpeech, yielding smaller absolute gains, but Latent Softmax consistently improves both interfaces.

Among Latent Softmax systems, LLM-P2G performs best on AISHELL-1 (5.26\%), whereas the projector performs best on LibriSpeech (3.63\%/9.14\%). Thus, neither interface dominates universally; the S2P improvements transfer to both.

\begin{table*}[t]
  \caption{Example subclass and major-class phoneme decoding on LibriSpeech. Subclass sequences are decoded from subclass posteriors; major-class sequences are decoded after marginalizing subclass posterior probabilities.}
  \vspace{-2mm}
  \label{tab:english_ipa}
  \centering
  \begin{tabular}{l l l}
    \toprule
    text & subclass phoneme sequence & major class phoneme sequence   \\
    \midrule
    where is that & w \`e  \textturnrrtail  \ \'i  \ z \dh  \ \ae  \ \textipa{t\textsuperscript{h}}  & w e \textturnrrtail  \ i z \dh  \ \ae  \ \textipa{t\textsuperscript{h}}   \\
    yes i know very well & j \`e s \'a \'i  n \'o  \'\textupsilon \ v \`\textschwa  \ \textturnrrtail  \ \'i w \`e  l & j e s a i n o \textupsilon  \ v \textschwa  \ \textturnrrtail  \ i w e l  \\
    what made the difference & w \`\textschwa \ \textipa{t\textsuperscript{h}} m \`e \=i d \dh \ \v{\textschwa} d \`i f \v{\textschwa} \textturnrrtail \ \v{\textschwa} n s & w \textschwa \ \textipa{t\textsuperscript{h}} m e i d \dh \ \textschwa \ d i f \textschwa \ \textturnrrtail \ \textschwa \ n s \\

    \bottomrule
  \end{tabular}
  \vspace{-3mm} 
\end{table*}

\begin{table}[t]
  \caption{LLM-P2G recognition error rates (\%) with major-class or subclass S2P phoneme inputs: CER on AISHELL-1 and WER on LibriSpeech (LS).}
  \vspace{-2mm}
  \label{tab:major_class_result}
  \centering
  \begin{tabular}{l|c|c|c}
    \toprule
    \textbf{S2P output} & \textbf{AISHELL-1} & \textbf{LS clean} & \textbf{LS other} \\
    & CER & WER & WER \\
    \midrule
    Major class & 5.31 & 4.10 & 10.19 \\
    Subclass & \textbf{5.26} & \textbf{3.91} & \textbf{9.89} \\
    \bottomrule
  \end{tabular}
  \vspace{-4mm}
\end{table}

\subsection{Fine-tuning on Code-switching Datasets}

The AISHELL-1 and LibriSpeech evaluations above pool Mandarin and English data, but each utterance is single-language rather than code-switched. We therefore further consider Mandarin--English code-switching, a more difficult setting where languages may mix within an utterance. Table~\ref{tab:code-switch} reports adaptation results obtained by fine-tuning the corresponding systems in Table~\ref{tab:exp}. We evaluate both LLM-P2G and projector interfaces on ASRU2019. For CS-Dialogue, the LLM-P2G training was unstable and is therefore omitted, as discussed below.

On ASRU2019, the two downstream interfaces behave differently. With LLM-P2G, standard softmax S2P gives a slightly lower MER than Latent Softmax (13.69 vs. 13.83), consistent with its lower non-tonal PER (3.23 vs. 3.83). With the projector interface, however, Latent Softmax reduces MER by 2.6\% relative to standard softmax (14.08 vs. 13.71), suggesting that the projector approach still benefits from the tone-aware representation learned by Latent Softmax.

On CS-Dialogue, LLM-P2G is harder to adapt, likely because the available code-switching data are limited and the decoded S2P phoneme sequences remain too noisy for training a robust text conversion model. We therefore report only projector-based MER for this dataset. In this setting, Latent Softmax improves both non-tonal PER and MER, achieving a 9.5\% relative MER reduction over standard softmax (15.53 vs. 14.06). Thus, the code-switching evidence supports projector-based MER improvements by Latent Softmax on both evaluated datasets; the LLM-P2G results do not establish a consistent advantage, which needs further investigation.

\section{Additional Analysis}

\subsection{Effect of S2P encoder improvements in LLM-ASR architectures}

Latent Softmax primarily improves S2P, so we examine how these gains affect downstream CER and WER. For LLM-P2G in Table~\ref{tab:exp}, downstream gains closely follow S2P PER reductions. Replacing standard softmax reduces PER by 0.17, 0.28, and 0.63 percentage points on AISHELL-1, test-clean, and test-other, respectively (tonal PER for Mandarin; non-tonal PER for English). Corresponding reductions are 0.37 points in AISHELL-1 CER and 0.38/0.81 points in LibriSpeech WER. The CER/PER and WER/PER ratios differ across evaluation conditions but remain similar between softmax methods within each condition. This suggests a relatively stable P2G conversion effect with respect to the emitted S2P symbols, while S2P quality improvements remain visible in downstream CER and WER.

For projector systems, the same S2P PER improvements reduce AISHELL-1 CER by 0.09 percentage points and LibriSpeech WER by 0.29 and 0.47 points on test-clean and test-other. These gains are smaller than those in the LLM-P2G setting. Presumably, this is because the projector interface does not directly rely on a discrete phoneme string; it uses continuous representations from the S2P encoder before large-language-model decoding. Nevertheless, consistent CER/WER gains across all three sets indicate that Latent Softmax improves upstream S2P representations, beyond merely changing the surface form of decoded phoneme symbols.

\subsection{Tonal representation analysis in non-tonal languages}
Table~\ref{tab:english_ipa} shows illustrative LibriSpeech examples decoded by the Latent Softmax S2P model. The subclass sequences are obtained by decoding the subclass posterior distribution. For major-class decoding, subclass posterior probabilities are first marginalized according to Eq.~(\ref{eq:marginal}), and beam search is then applied to the resulting major-class distribution.

The examples show that, even without explicit English tone labels, the model assigns tone-like subclass marks to English vowels. After marginalization, these marks collapse back to the expected major-class phoneme sequence. We treat this observation as qualitative evidence that the latent tone-marked vowel subclass space captures prosodic variation useful for sharing with tonal-language data. It should not be interpreted as a claim that the predicted English tone marks are linguistically ground-truth tones. English still exhibits pitch variation in speech production, but such variation is non-contrastive and therefore unmarked in the phonetic transcription. From the model's viewpoint, an English vowel token can nevertheless instantiate one latent tone-marked vowel subclass inside the corresponding major class. A preliminary inspection of the utterances in Table~\ref{tab:english_ipa} suggests qualitative agreement between the directions of the predicted subclass tone marks and their F0 contours. This is qualitative evidence only, not corpus-level validation.

\subsection{Subclass versus major-class phoneme input in LLM-P2G}
Table~\ref{tab:major_class_result} compares LLM-P2G inputs from major-class and subclass S2P outputs. Subclass inputs reduce AISHELL-1 CER from 5.31\% to 5.26\% and LibriSpeech WER from 4.10\% to 3.91\% on test-clean and from 10.19\% to 9.89\% on test-other. The absolute gains range from 0.05 to 0.30 percentage points. This suggests that the subclass representation contains useful fine-grained information for the P2G model, including in the non-tonal English setting. At the same time, the gains are modest, so the result should be understood as supporting evidence rather than a standalone proof of tonal correctness.

\section{Conclusion}

This paper studied data-efficient combined training of tonal and non-tonal languages for phoneme-based multilingual ASR. Latent Softmax keeps tone-marked vowels as subclasses, base vowels as major classes, and consonants and the blank as singleton labels. When only a non-tonal base-vowel label is available, its subclass is latent and marginalized. Non-tonal data thus update the tone-marked subclass space without erasing tone distinctions, improving sharing over standard softmax. On pooled AISHELL-1 Mandarin and LibriSpeech English, S2P PER gains transfer consistently to both LLM-P2G and projector-based ASR. For Mandarin--English code-switching, Latent Softmax improves projector-based MER on ASRU2019 and CS-Dialogue, whereas LLM-P2G results remain inconclusive. Future work can extend the method to more tonal languages and larger multilingual training data, and quantify the relation between latent subclasses and F0 at corpus scale.

\clearpage
\section*{GENERATIVE AI USE DISCLOSURE}
Generative AI tools are used in this work only for language editing, polishing, and formatting of the manuscript. They are not used to generate any core content, research ideas, experimental designs, results, or major textual parts of the paper. All scientific contributions, including model design, experiments, analysis, and conclusions, are completed by the authors.

\bibliographystyle{IEEEtran}
\bibliography{mybib}
\clearpage
\appendix[Proof of Equations for CTC Likelihood with Latent Softmax]
\phantomsection\label{app:latent_softmax_ctc_proof}
\begingroup
\color{black}

\subsection{Label inventories and compatible subclasses}
\label{app:inventory}

This appendix derives the aggregated (marginal) frame posterior in Eq.~(\ref{eq:marginal}) and the resulting CTC likelihood in Eq.~(\ref{eq:ctc_latent}). Let $\blank$ denote the CTC blank. Recall that $\mathcal{Z}$ is the subclass output inventory of the neural network and includes $\blank$. Define the nonblank subclass inventory as $\mathcal{Z}_{\mathrm{nb}}=\mathcal{Z}\setminus\{\blank\}$. Let $\mathcal{L}_{\mathrm{nb}}$ denote the set of nonblank supervised labels that can appear in training target sequences. It is decomposed as
\begin{equation}
\mathcal{L}_{\mathrm{nb}}
=\mathcal{Z}_{\mathrm{nb}}\cup\mathcal{M},
\qquad
\mathcal{Z}_{\mathrm{nb}}\cap\mathcal{M}=\varnothing.
\label{eq:proof_label_sets}
\end{equation}
Here, $\mathcal{Z}_{\mathrm{nb}}$ contains the fine-grained subclass labels, including tone-marked vowels, consonants, and base vowels that have no tone-marked counterparts (e.g., the English phoneme /\textipa{A}/). The set $\mathcal{M}$ contains the base-vowel major-class labels whose tone-marked subclasses are included in $\mathcal{Z}_{\mathrm{nb}}$, such as /i/, /\textipa{@}/, /e/, /a/, /o/, /\textipa{U}/, and /u/. Labels in $\mathcal{M}$ have no independent output logits.

For each possible alignment label $l\in\mathcal{L}_{\mathrm{nb}}\cup\{\blank\}$, let $B(l)\subseteq\mathcal{Z}$ denote its compatible-subclass set, as defined in Eq.~(\ref{eq:Bset}), with $B(\blank)=\{\blank\}$. Therefore,
\begin{equation}
\lvert B(l)\rvert=
\begin{cases}
1, & l\in\mathcal{Z}_{\mathrm{nb}}\cup\{\blank\},\\
\geq 2, & l\in\mathcal{M}.
\end{cases}
\label{eq:proof_B_size}
\end{equation}

Let $\x=(x_1,\ldots,x_T)$ be an acoustic observation sequence and $\y=(y_1,\ldots,y_N)$ its target phoneme sequence, where each $y_n\in\mathcal{L}_{\mathrm{nb}}$ is a supervised-label occurrence and its value is a supervised label. This formulation also accommodates code-switched utterances: the supervised labels at different positions in $\y$ may belong to either $\mathcal{Z}_{\mathrm{nb}}$ or $\mathcal{M}$.

CTC introduces a frame-level alignment $\bpi=(\pi_1,\ldots,\pi_T)$, where $\pi_t\in\mathcal{L}_{\mathrm{nb}}\cup\{\blank\}$ and $\bpi\in\mathcal{A}(\y)$. The network softmax directly provides $p_t(z)$ for every $z\in\mathcal{Z}$, but it provides no separate posterior for a major-class label in $\mathcal{M}$. Thus, $p_t$ is the normalized frame-level distribution over $\mathcal{Z}$; each major-class quantity $q_t(l)$ is a marginal event probability derived from $p_t$, not a separate softmax output.

\newpage
\subsection{Marginalization for a fixed alignment}
\label{app:alignment}

When $\pi_t$ is a major-class label, the compatible fine-grained subclass $z_t$ is unobserved. Let $\z=(z_1,\ldots,z_T)$ with $z_t\in\mathcal{Z}$. For a fixed CTC alignment $\bpi$, the indicator $\mathbb{I}\{z_t\in B(\pi_t)\}$ selects subclass paths compatible with that alignment. Marginalizing over these paths gives
\begin{align}
P(\bpi\mid\x)
&=\sum_{\z\in\mathcal{Z}^{T}}
\prod_{t=1}^{T}
\left[p_t(z_t)\,\mathbb{I}\{z_t\in B(\pi_t)\}\right]\nonumber\\
&=\prod_{t=1}^{T}
\sum_{z_t\in B(\pi_t)}p_t(z_t).
\label{eq:proof_alignment}
\end{align}
The first summation is equivalently a sum over the Cartesian-product set $B(\pi_1)\times\cdots\times B(\pi_T)$. The second equality follows from the frame-wise factorization of the subclass-path probability, which allows the compatible subclass at each frame to be marginalized independently.

\subsection{CTC sequence likelihood}
\label{app:sequence}

Consequently, summing over all valid CTC alignments yields
\begin{align}
P(\y\mid\x)
&=\sum_{\bpi\in\mathcal{A}(\y)}P(\bpi\mid\x)\nonumber\\
&=\sum_{\bpi\in\mathcal{A}(\y)}
\prod_{t=1}^{T}
\sum_{z_t\in B(\pi_t)}p_t(z_t).
\label{eq:proof_latent_ctc}
\end{align}
Defining
\begin{equation}
q_t(l)=\sum_{z\in B(l)}p_t(z),
\qquad
l\in\mathcal{L}_{\mathrm{nb}}\cup\{\blank\},
\label{eq:proof_q}
\end{equation}
recovers Eq.~(\ref{eq:marginal}) and transforms Eq.~(\ref{eq:proof_latent_ctc}) exactly into Eq.~(\ref{eq:ctc_latent}). Thus, Latent Softmax preserves the standard CTC likelihood form and forward--backward recursion while marginalizing the unobserved subclasses compatible with a major-class label.

\endgroup
\end{document}